\documentclass{article}

 \usepackage{graphicx}
 \usepackage{wrapfig}
 \usepackage{algorithm}
 \usepackage{algpseudocode}
 \usepackage{titlesec}[compact]

\usepackage[final]{neurips_2025}
\usepackage[utf8]{inputenc} % allow utf-8 input
\usepackage[T1]{fontenc}    % use 8-bit T1 fonts
\usepackage{hyperref}       % hyperlinks
\usepackage{url}            % simple URL typesetting
\usepackage{booktabs}       % professional-quality tables
\usepackage{amsfonts}       % blackboard math symbols
\usepackage{nicefrac}       % compact symbols for 1/2, etc.
\usepackage{microtype}      % microtypography
\usepackage{xcolor}         % colors

\title{GraphFaaS: Serverless GNN Inference for Burst-Resilient, Real-Time Intrusion Detection}

\author{%
  Lingzhi Wang\\
  Northwestern University\\
  \And
  Vinod Yegneswaran \\
  SRI International \\
  \And
  Xinyi Shi \\
  Northwestern University \\
  \AND
  Ziyu Li \\
  Northwestern University \\
  \And
  Ashish Gehani \\
  SRI International\\
  \And
  Yan Chen \\
  Northwestern University \\
}

\begin{document}

\maketitle

\begin{abstract}
Provenance-based intrusion detection is an increasingly popular application of graphical machine learning in cybersecurity, where system activities are modeled as provenance graphs to capture causality and correlations among potentially malicious actions. Graph Neural Networks (GNNs) have demonstrated strong performance in this setting. However, traditional statically-provisioned GNN inference architectures fall short in meeting two crucial demands of intrusion detection: (1) maintaining consistently low detection latency, and (2) handling highly irregular and bursty workloads.
To holistically address these challenges, we present GraphFaaS, a serverless architecture tailored for GNN-based intrusion detection.
GraphFaaS leverages the elasticity and agility of serverless computing to dynamically scale the GNN inference pipeline.
We parallelize and adapt GNN workflows to a serverless environment, ensuring that the system can respond in real time to fluctuating workloads. By decoupling compute resources from static provisioning, GraphFaaS delivers stable inference latency, which is critical for dependable intrusion detection and timely incident response in cybersecurity operations.
Preliminary evaluation shows GraphFaaS reduces average detection latency by 85\% and coefficient of variation (CV) by 64\% compared to the baseline.
\end{abstract}

\section{Introduction}
\vspace{-0.1in}
\begin{wrapfigure}{r}{0.37\textwidth}
  \centering
  \includegraphics[width=0.35\textwidth]{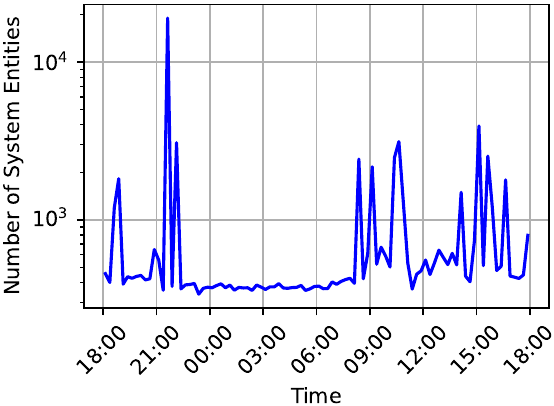}
  \caption{Fluctuations in detection workload over time in the DARPA TC dataset~\cite{noauthor_transparent_nodate}}
  \label{fig:wrap-node-count}
  \vspace{-0.1in}
\end{wrapfigure}

Graph-based intrusion detection has emerged as a critical application of graph analysis in cybersecurity.
In this paradigm, system behaviors recorded in audit logs are transformed into provenance graphs, 
where nodes represent system entities (e.g., processes, files, network nodes), and 
edges represent system-level events (e.g., file read/write, process creation).
These graphs provide a natural way to model the causal and temporal relationships underlying cyberattacks and anomalous behavior.
Recently, Graph Neural Networks (GNNs) have demonstrated strong promise in this domain by learning
complex patterns that enable context-aware intrusion detection. 
However, existing GNN inference architectures struggle to scale in this setting, as they must (1) maintain stable and low detection latency, while also (2) handling inherently irregular and bursty workloads (i.e., workloads with sudden surges in demand).
First, in intrusion detection systems, it is particularly crucial for GNN inference to maintain stability and low latency.
If the inference time is too long, there is a high risk of missing the window to mitigate real attacks, potentially leading to irreversible consequences.
Therefore, ensuring fast and consistent inference performance is essential for effective threat response.
Second, such systems often face unstable and bursty workloads (Figure~\ref{fig:wrap-node-count}), while malicious activities account for only a small fraction of overall network behavior, leading to highly unpredictable and sporadic patterns in the analysis workload.

% \begin{figure}
%     \centering
%     \includegraphics[width=0.5\linewidth]{figures/workload.pdf}
%     \caption{Caption}
%     \label{fig:placeholder}
% \end{figure}

Serverless design offers advantages in handling bursty workload and maintaining consistent low detection latency.
A serverless system is a cloud model where code runs on-demand without managing servers.
It provides a cost-efficient pay-per-use model with no charges for idle resources, while scaling instantly and seamlessly to handle traffic spikes.
Additionally, it enables faster development through an event-driven model and ensures finer-grained resource utilization, making it especially well-suited for workloads that are intermittent, unpredictable, or highly event-driven.

\noindent{\bf Contributions.}  This paper introduces GraphFaaS, a serverless architecture for low-latency, scalable GNN-based intrusion detection. Our contributions include: ($i$) A {\bf provenance-aware graph construction pipeline} that avoids redundant computation by leveraging temporal locality in system ls and applying two-stage filtering, based on structural proximity and frequency, to isolate relevant subgraphs.  ($ii$) A {\bf serverless node embedding layer}, which parallelizes node-level feature transformations across dynamic workloads with a feature-length-aware execution strategy to ensure each unit completes within a fixed latency bound while minimizing overhead. ($iii$) A {\bf scalable GNN inference mechanism}, adapted for serverless execution, which partitions large provenance graphs into balanced subgraphs using a greedy best-fit algorithm and a vertical-scaling fallback to handle extreme cases where graph fragments exceed preset limits, ($iv$) An {\bf end-to-end implementation} of GraphFaaS on the OpenFaaS platform, evaluated on the DARPA TC dataset~\cite{noauthor_transparent_nodate}. Our system achieves a 6.7x reduction in average detection latency and a 64\% reduction in coefficient of variation (CV) compared to a state-of-the-art baseline, while maintaining equivalent detection accuracy.

%We leverage the benefits of serverless design to tackle the challenges of deploying GNN inference in provenance-based intrusion detection systems.
%We decompose the workflow of GNN-based intrusion detection into serverless functions and parallelize the workload into subtasks, ensuring that each subtask remains below a predefined threshold.
%With the automated scaling capability of serverless functions, the number of function instances dynamically adjusts to match the changing number of subtasks.
%When the workload is high, additional function instances are provisioned to maintain low latency; when the workload is low, unnecessary instances are terminated to minimize cost.
%GraphFaaS reduces average detection latency by 85\% and variability (COV) by 64\% compared to the baseline.
% We first introduce the typical workflow of existing GNN-based PIDS in \S\ref{sec:typical_workflow}.
% In \S\ref{sec:design}, we present the proposed serverless GNN inference architecture and demonstrate how it is applied to develop a serverless GNN-based PIDS.

\vspace{-0.1in}
\section{Background and Related Work}
\vspace{-0.1in}
\textbf{Provenance-based Intrusion Detection.}
Intrusion detection is the process of monitoring and analyzing computer systems or networks to identify unauthorized access, malicious activity, or policy violations.
In provenance-based intrusion detection systems (PIDS), system behaviors are modeled as directed acyclic graphs (provenance graphs), where nodes represent system entities (e.g., processes, files, sockets, pipes, memory objects) and edges capture interactions between these entities (e.g., reading a file or connecting to a remote host).
The provenance graphs evolve dynamically as the system operates, forming a temporal record that encodes the causal relationships among system entities, which is then used to correlate the suspicious traces in the cyberattacks.
In recent years, graph machine learning methods, particularly GNNs, have become a prominent approach for detecting advanced persistent threats (APTs) from provenance graphs~\cite{Abrar25,Bilot25,rehman2024flash,cheng2024kairos,jiang2025orthrus}.
Typically, continuous log streams are transformed into fixed-duration graphs, which are then processed by GNN models.
Most PIDSs are trained to learn the benign graph patterns.
During inference, the GNN model takes a graph snippet as input.
The resulting node and graph embeddings are then analyzed to detect deviations from benign graph patterns.
% Detection is typically carried out using clustering algorithms, fixed thresholds, or outlier detection techniques to flag anomalous nodes, edges, or subgraphs.

\textbf{Serverless for GNN.}
Recent efforts have tackled the challenges of efficient GNN execution from both hardware and system perspectives.
GNNAdvisor~\cite{wang2021gnnadvisor} optimizes GPU utilization by tailoring runtime behaviors based on GNN model and workload characteristics, but focuses mainly on training/inference under static resources.
$\lambda$ Grapher~\cite{hu2024lambdagrapher} enables efficient GNN serving by exploiting request-level graph locality and fine-grained resource control.
Dorylus~\cite{thorpe2021dorylus} leverages serverless threads with CPU servers for scalable GNN training, achieving cost-efficiency via computation separation, yet it does not address serving workloads.
Fograph~\cite{zeng2022fograph} targets low-latency GNN inference for IoT by using fog nodes close to data sources, mitigating cloud communication overhead.
However, it relies on specialized edge deployments rather than general-purpose serverless infrastructures.
While some works explore migrating graph processing to FaaS platforms, they suffer from poor scalability due to high communication overhead and coarse-grained execution~\cite{toader2019graphless}.
Unlike prior systems that focus on static provisioning, training performance, or specialized edge deployments, GraphFaaS uniquely enables low-latency, scalable GNN inference for intrusion detection by combining fine-grained graph filtering, adaptive partitioning, and dynamic resource scaling within a serverless framework.

%{]\bf VY:  NEED A LINE HERE THAT EMPHASIZES HOW GraphFaaS is different.}]  GraphFaaS distinguishes itself from prior systems by ....}

\vspace{-0.1in}
\section{Typical Detection Workflow of GNN-based PIDS}\label{sec:typical_workflow}
\vspace{-0.1in}
\textbf{Log processing and graph construction.}
The source of intrusion detection is the log stream,i.e., system log data arranged in chronological order.
Graph construction extracts system entities and events from the log stream, representing them as nodes and edges in the provenance graph, and stores them in a graph database.

\textbf{Node-level embedding.}\label{sec:node-level-embedding}
The second step is node-level embedding, also known as featurization~\cite{Bilot25}, which transforms the static attributes of each node into numerical vectors. These vectors act as the initial node representations for the GNN. In provenance graphs, node attributes are typically represented as textual data, for example, process names and command lines for process nodes, file paths for file nodes, and IP addresses and ports for network nodes.
To embed these textual features into a vector space, common techniques such as word2vec and doc2vec are often employed.

\textbf{GNN.}
The node-level embeddings provide the initial representations for GNN.
During message passing, the GNN propagates and aggregates node features across its neighbors. The scope of this propagation, i.e., the receptive field, is determined by the number of GNN layers. After propagation, each node is assigned an updated vector representation.
To train the GNN parameters, typical optimization objectives include node classification, edge type prediction, or graph reconstruction.

\textbf{Result Aggregation.}
GNN generates an embedding for each node. For an intrusion detection system, these embeddings are further analyzed to determine whether an intrusion has occurred. Common approaches in existing work include applying clustering methods, fixed thresholds, or outlier detection techniques to identify anomalous nodes.

\vspace{-0.1in}
\section{Serverless GNN-based Intrusion Detection Architecture}\label{sec:design}
\begin{figure}
    \centering
    \includegraphics[width=0.85\linewidth]{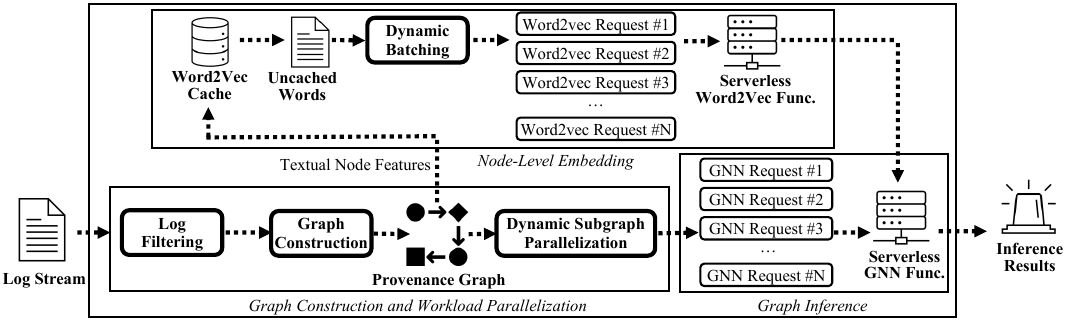}
    \vspace{-0.1in}
    \caption{Overview of GraphFaaS Framework with three components: (i) graph construction, (ii) serverless node-level embedding, and (iii) serverless GNN inference.}
    \label{fig:flowchart}
    \vspace{-0.2in}
\end{figure}

\vspace{-0.1in}
%We propose leveraging serverless architecture for real-time, burst-resilient GNN-based intrusion detection that offers stable detection latency.
As shown in Figure~\ref{fig:flowchart}, the GraphFaaS framework is composed of three main components: (i) graph construction, (ii) serverless node-level embedding, and (iii) serverless GNN inference.

\textbf{Graph Construction.}
We start by constructing the provenance graph from the log stream.
Although the provenance graph can be very large, we find that most of its structure stays the same between two detection intervals.
To take advantage of this, we use log filtering: instead of reprocessing the entire graph, we focus only on the parts that have changed.
Specifically, we keep nodes that are within a 2K-hop distance of the active nodes (where the GNN has K layers), which avoids redundant computation.
We can also apply frequency-based filtering~\cite{dong2023distdet,hassan2019nodoze}: we keep only the edges, nodes, and their 2K-hop neighborhoods that occur infrequently in the training data, since rare patterns are usually more important for detection.
Finally, the filtered subgraphs are split into subtasks and run in parallel, allowing us to fully leverage serverless auto-scaling.

\textbf{Serverless Node-level Embedding.}
Our goal is to ensure consistent and low inference latency under bursty workloads.
The core idea is to divide the workload into small, parallel execution units, each requiring only a short processing time below a predefined threshold.
As the workload changes, the number of execution units adjusts accordingly.
Leveraging the automatic scaling capability of the serverless architecture, computing resources are allocated adaptively to match demand.
This design guarantees that all execution units complete within the time threshold, regardless of their number.
For instance, when a large workload arrives, it is split into more execution units, and the serverless platform automatically scales up by launching additional instances to handle them.
Conversely, when the workload is low, the platform scales down to release resources and avoid waste.
GNN inference typically consists of two stages: node-level embedding and message passing with aggregation.
We present how each of these stages can be adapted to a serverless architecture.

As introduced in \S\ref{sec:node-level-embedding}, node-level embedding transforms node attributes into vectors that serve as the initial representations for GNNs.
Because static node attributes are independent of one another, the embedding process is naturally parallelizable across nodes.
We implement a serverless function that takes a node attribute as input and returns its vector embedding.
Since node attributes are typically represented as textual strings, and execution time for most embedding methods depends on string length, we partition execution units according to feature length.
Shorter strings are grouped into a single execution unit, while longer strings are processed separately.
This strategy ensures that each execution unit completes within the designated time threshold while also reducing the overhead of excessive parallelization, such as network transmission and packet processing costs.

{\bf GNN Inference.} We implement a serverless function that takes the initial node embeddings and edges as input and outputs the updated node vectors.
Similar to the embedding stage, the workload is divided into execution units so that the serverless platform can scale automatically to handle bursty workloads.
% The latency of GNN inference depends on both the model architecture and the size of the input graph.
Since the model remains fixed during detection, inference latency is primarily determined by the graph size.
To efficiently process large graphs, we partition them into subgraphs whose sizes are close to a predefined threshold.
This partitioning strategy needs to address the key challenge of workload balancing, specifically, how to partition or pack the graph such that each subgraph stays within the desired size limit while minimizing the total number of subgraphs, thereby avoiding excessive waste of computational resources.
We design a greedy best-fit algorithm to partition the graph (See Appendix\ref{sec:greedy-partition} for more details). 
Due to the dependency explosion problem in provenance graphs, even the smallest subgraph (a central node and its k-hop neighbors) may exceed the preset size limit.
In such cases, vertical scaling is triggered, allocating more compute resources (e.g., CPU cores, memory) to each serverless function instance, instead of spawning more instances.

\vspace{-0.1in}
\section{Preliminary Results}
\vspace{-0.05in}
\vspace{-0.1in}
We implemented GraphFaas in Python based on OpenFaaS, an open-source serverless platform.
We compared GraphFaas with Flash~\cite{rehman2024flash}, a state-of-the-art PIDS.
To ensure a fair comparison, we reimplemented Flash to adapt it to a server-client detection model, where data is sent from the monitored machine to a detection server for analysis.
The only difference is that we used a Docker container to simulate a statically provisioned detection environment for Flash, whereas GraphFaas was deployed using a serverless architecture based on OpenFaaS.
All other experimental conditions were kept identical between the two systems.
We used a widely adopted dataset from DARPA TC Engagement 3~\cite{noauthor_transparent_nodate}, which contains 11 days of audit logs and includes four attack campaigns.

\begin{figure}
    \centering
    \includegraphics[width=0.9\linewidth]{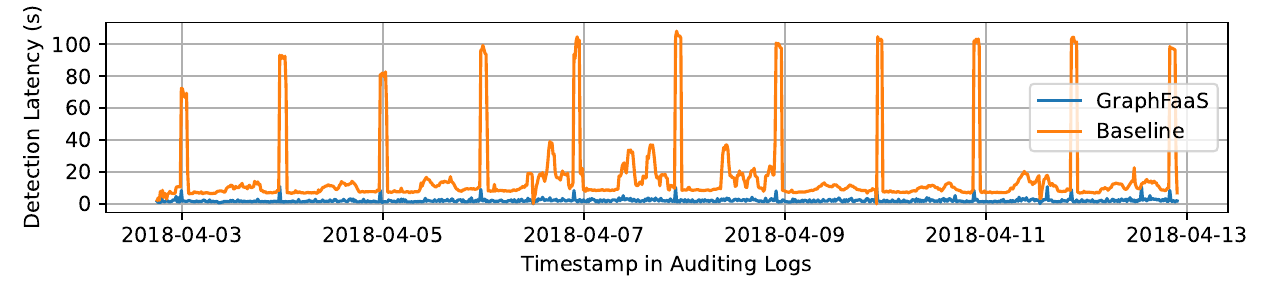}
   \vspace{-0.2in}
    \caption{Detection latency over time. GraphFaaS: mean = 2.10, STD (standard deviation) = 1.09, CV (coefficient of variation) = 0.52;
    Baseline: mean = 14.16, STD = 4498.92, CV = 1.46.}
    \label{fig:latency-compare}
    \vspace{-0.2in}
\end{figure}

We verified that GraphFaaS achieves the same detection accuracy as Flash, which is expected since the underlying detection models of Flash remain unchanged.
Therefore, the detection results are identical.
However, GraphFaaS significantly outperforms Flash in detection latency.
As shown in Figure~\ref{fig:latency-compare}, on a 11-day dataset, it achieves an average latency of 2.1 seconds, compared to 14.16 seconds for the baseline.
In addition, GraphFaaS is more resilient to bursty traffic.
It maintains low latency even during sudden surges in workload, as evidenced by its lower standard deviation and coefficient of variation.
While occasional latency spikes occur, the delay never exceeds 10 seconds.

\vspace{-0.1in}
\section{Conclusion and Future Work}
\vspace{-0.05in}
\vspace{-0.1in}
GraphFaaS leverages the elasticity and agility of a serverless design to ensure consistently low intrusion detection latency.
% Our results demonstrate that, compared with statically provisioned architectures, GraphFaaS maintains stable detection latency under bursty workloads.
Several challenges remain for future work.
First, provenance graphs are particularly vulnerable to the dependency explosion problem~\cite{hossain2020combating,wang2024incorporating}, where certain super-nodes dominate the graph.
Even with workload parallelization, these super-nodes remain a major latency bottleneck.
Second, during cyberattacks, the scale of the provenance graph can fluctuate dramatically.
To effectively capture long-range dependencies while preserving low latency, it is necessary to dynamically adjust the number of GNN layers and invoke the appropriate serverless functions on demand.  
We believe GraphFaaS can serve as the basis to address these issues with serverless design.  Finally, we plan to evaluate GraphFaaS against a broader class of PIDs and across more datasets. 

\ack
This work was supported in part by the National Science Foundation (NSF) under Grant No. CNS-2229455.
The authors gratefully acknowledge the support of the NSF, which made this research possible.

\bibliographystyle{plain}
\bibliography{ProvFaas}

%\section*{References}
% References follow the acknowledgments in the camera-ready paper. Use unnumbered first-level heading for
% the references. Any choice of citation style is acceptable as long as you are
% consistent. It is permissible to reduce the font size to \verb+small+ (9 point)
% when listing the references.
% Note that the Reference section does not count towards the page limit.
% \medskip

% {
% \small

% [1] Alexander, J.A.\ \& Mozer, M.C.\ (1995) Template-based algorithms for
% connectionist rule extraction. In G.\ Tesauro, D.S.\ Touretzky and T.K.\ Leen
% (eds.), {\it Advances in Neural Information Processing Systems 7},
% pp.\ 609--616. Cambridge, MA: MIT Press.

% [2] Bower, J.M.\ \& Beeman, D.\ (1995) {\it The Book of GENESIS: Exploring
%   Realistic Neural Models with the GEneral NEural SImulation System.}  New York:
% TELOS/Springer--Verlag.

% [3] Hasselmo, M.E., Schnell, E.\ \& Barkai, E.\ (1995) Dynamics of learning and
% recall at excitatory recurrent synapses and cholinergic modulation in rat
% hippocampal region CA3. {\it Journal of Neuroscience} {\bf 15}(7):5249-5262.
% }

%%%%%%%%%%%%%%%%%%%%%%%%%%%%%%%%%%%%%%%%%%%%%%%%%%%%%%%%%%%%

\appendix
\section{Algorithm for Subgraph Partition \& Packing}\label{sec:greedy-partition}

The goal of subgraph partitioning \& packing is to divide a large graph into several smaller subgraphs such that the size of the K-hop neighborhood of each subgraph remains within a predefined threshold.
At the same time, we aim to avoid excessive fragmentation by merging K-hop neighborhoods that share overlapping regions.
In other words, we seek a partitioning of the original graph where each subgraph’s K-hop neighborhood is bounded in size, while also minimizing the total number of subgraphs.
For each cluster \(k\), we maintain its current node set \(U_k = N_2(X_k)\), represented as a bitset, and its edge count \(f_k = |E_G[U_k]|\). When a new vertex \(v\) is considered for insertion, we compute the increase in edge count as:
\[
\Delta_k = \left|E_G[U_k \cup B_2(v)]\right| - f_k.
\]
If \(\Delta_k\) does not exceed the remaining capacity of cluster \(k\), the vertex is inserted into that cluster. Otherwise, a new cluster is created.

To minimize the total number of clusters used, we adopt the \emph{Best-Fit} strategy, which attempts to pack the vertex into the cluster where it fits most tightly. A simpler alternative is the \emph{First-Fit} strategy.

\newpage

\begin{algorithm}[H]
\caption{BinPack\_FFD (First-Fit Decreasing Bin Packing) for Graph Partitioning}
\begin{algorithmic}[1]
\Require List of neighborhoods \texttt{neighbors}, capacity \texttt{capacity}, strict flag \texttt{strict}
\Ensure List of bins and remaining capacities

\State Sort \texttt{neighbors} by number of edges in descending order
\State Initialize \texttt{bins} as empty list
\State Initialize \texttt{remain} as empty list
\State Initialize \texttt{subgraph} as empty list of edge sets

\For{each \texttt{(idx, neighborhood)} in sorted items}
    \State \texttt{placed} $\gets$ \texttt{False}
    \State \texttt{edge\_set\_to\_add} $\gets$ set of edges in \texttt{neighborhood}
    
    \For{$b = 0$ to $|\texttt{bins}| - 1$}
        \State \texttt{margin} $\gets$ size of \texttt{edge\_set\_to\_add} minus overlapping edges with \texttt{subgraph[b]}
        \If{\texttt{strict} is True}
            \If{\texttt{margin} $<$ \texttt{remain[b]}}
                \State Add \texttt{idx} to \texttt{bins[b]}
                \State Update \texttt{subgraph[b]} and \texttt{remain[b]}
                \State \texttt{placed} $\gets$ \texttt{True}, \textbf{break}
            \EndIf
        \Else
            \If{\texttt{margin} $\leq$ \texttt{remain[b]}}
                \State Add \texttt{idx} to \texttt{bins[b]}
                \State Update \texttt{subgraph[b]} and \texttt{remain[b]}
                \State \texttt{placed} $\gets$ \texttt{True}, \textbf{break}
            \EndIf
        \EndIf
    \EndFor

    \If{\texttt{placed} is \texttt{False}}
        \State Create new bin with \texttt{idx}
        \State Add \texttt{edge\_set\_to\_add} to \texttt{subgraph}
        \State Append \texttt{capacity - |edge\_set\_to\_add|} to \texttt{remain}
    \EndIf
\EndFor

\State \Return \texttt{bins}, \texttt{remain}
\end{algorithmic}
\end{algorithm}

\end{document}